\documentclass[twocolumn,superscriptaddress,showpacs, longbibliography, aps, prb]{revtex4-1}
\ifx\pdfoutput\undefined
\usepackage[dvipdfmx]{graphicx}
\usepackage[dvipdfmx]{hyperref}
\usepackage[dvipdfmx]{color}
\else
\usepackage{graphicx}
\usepackage{hyperref}
\usepackage{color}
\fi
\usepackage{amsmath, amssymb}
\hypersetup{colorlinks=true,breaklinks,linkcolor=blue,urlcolor=blue,citecolor=blue}
\usepackage{dcolumn}
\usepackage{bm}
\usepackage{xspace}
\usepackage{physics}

\def\vec#1{\boldsymbol #1}

\usepackage{ulem}

\newcommand{\HPhi}{\mathcal{H}\Phi}

\def\vec#1{\boldsymbol #1}
\usepackage{natbib}

\begin{document}

\preprint{}

\title{Interedge spin resonance in the Kitaev quantum spin liquid}

\author{Takahiro Misawa}
\affiliation{Beijing Academy of Quantum Information Sciences, Haidian District, Beijing 100193, China}
\author{Joji Nasu}
\affiliation{Department of Physics, Tohoku University, Sendai 980-8578, Japan}
\author{Yukitoshi Motome}
\affiliation{Department of Applied Physics, University of Tokyo, Tokyo 113-8656, Japan}

\date{\today}

\begin{abstract}
The Kitaev model offers a platform for quantum spin liquids (QSLs) 
with fractional excitations, itinerant Majorana fermions and localized fluxes. 
Since these fractional excitations could be utilized for quantum computing, 
how to create, observe, and control them  through the spin degree of freedom is a central issue. 
Here, we study dynamical spin transport in a wide range of frequency for the Kitaev-Heisenberg model, 
by applying an AC magnetic field to 
an edge of the system.
We find that, in the Kitaev QSL phase, spin polarizations 
at the other edge are resonantly induced in a specific spin component, 
even though the static spin correlations are vanishingly small. 
This interedge spin resonance appears around the input frequency over the broad frequency range.
Comparing with the dynamical spin correlations,
we clarify that the resonance is governed by the itinerant Majorana fermions 
with a broad continuum excitation spectrum, which can 
propagate over long distances, although it 
vanishes for the pure Kitaev model because of accidental degeneracy and requires weak Heisenberg interactions.
We also find that the spin polarizations in the other spin components 
are weakly induced at an almost constant frequency close to the excitation gap of the localized fluxes, 
irrespective of the input frequency. 
These results demonstrate that the dynamical spin transport is a 
powerful probe of the fractional excitations in the Kitaev QSL. 
Possible experimental realization of the interedge spin resonance is discussed. 
\end{abstract}

\maketitle
\section{Introduction}
\label{sec:intro}
Exotic quasiparticles emerging in solids have attracted
much interest from both 
fundamental physics and industry applications.
A prominent example is Majorana particles --- charge-neutral spin-$1/2$ particles 
that are their own antiparticles~\cite{Majorana_1937,Wilczek_NPhys2009}. 
While they usually behave as fermions, 
in some two-dimensional cases they can be regarded as 
anyons that obey neither 
Fermi-Dirac nor Bose-Einstein statistics.
Such Majorana particles have been intensively studied for applications to quantum computing by 
using the anyonic nature~\cite{Kitaev_AP2003,Freedman_BA2003}. 

The Kitaev model on a honeycomb lattice 
offers an ideal platform for realizing the Majorana particles~\cite{Kitaev_ANP2006}. 
It is an exactly solvable model whose ground
state is a quantum spin liquid (QSL).
The Kitaev QSL 
hosts two types of emergent quasiparticles from the fractionalization of spins: itinerant Majorana fermions and localized fluxes. 
These quasiparticles turn into Abelian anyons when the interactions between spins are 
largely anisotropic, or non-Abelian anyons when an external magnetic field is applied in the nearly isotropic cases~\cite{Kitaev_AP2003,Kitaev_ANP2006}.
The Kitaev model could be realized 
in Mott insulators with the strong spin-orbit coupling~\cite{Jackeli_PRL2009}, such as 
Na$_{2}$IrO$_{3}$~\cite{Chaloupka_PRL2010} 
and $\alpha$-RuCl$_{3}$~\cite{Yadav_2016SciRep}. 
Detailed comparisons between experimental results 
and theoretical calculations have revealed 
fingerprints of the Majorana particles in 
{such} candidate materials; 
for a review, see Ref.~\onlinecite{Motome_JPSJ2020}.
Among them, the discovery of the half-quantized thermal Hall 
effect in $\alpha$-RuCl$_{3}$  
{was} ground breaking, providing  
direct evidence 
for the Majorana particles~\cite{Kasahara_Nature2018,Yamashita_PRB2020,Yokoi_Science2021}, 
while it is still under debate
~\cite{Bruin_NaturePhys2022,Czajka_NatureMat2022,Lefrancois_PRX2022}. 

Since the Majorana fermions and the fluxes 
in the Kitaev QSL are generated by the fractionalization of spins, 
they are  quantum entangled and inherently nonlocal. 
Indeed, the spatial correlations between the Majorana fermions 
are long-range with power-law decay~\cite{Willans_PRB2011,Koga_PRB2021}, although the spin correlations 
are short-ranged and vanish 
beyond nearest-neighbor 
sites~\cite{Baskaran_PRL2007}.
Furthermore, in the presence of defects or edges,
the spin correlations can be long-range due to 
low-energy excitations around the defects or edges
~\cite{Willans_PRL2010,Willans_PRB2011,Koga_PRB2021,Takikawa_PRB2022,Takahashi_arXiv2022}.

By exploiting such nonlocal nature, it was recently shown that the itinerant Majorana fermions can contribute to long-range spin transport from an edge of the system~\cite{Minakawa_PRL2020,Taguchi_JPS2023}. 
The previous study has focused only on low-energy properties, such as the velocity of the spin propagation determined by the slope of the gapless Majorana dispersion. 
However, the spin dynamics in the wider range of frequency is expected to offer more important insights into the two types of fractional quasiparticles with distinct excitation spectra. 
Such comprehensive study of nonlocal spin dynamics would also be a crucial step toward quantum computing, by elucidating 
how to control and probe the fractional quasiparticles
via the spin degree of freedom.

In this paper, in order to deepen the 
understanding of the relationship between the fractional quasiparticles and 
the spin degree of freedom, we study
nonlocal spin dynamics in the Kitaev QSL in the wide frequency range.
Applying a local AC magnetic field to one edge of the system, we investigate
how the spin excitations are excited and propagate 
to the other edge.
At the edges of the Kitaev model, 
it is known that local magnetic fields excite the fluxes accompanied by
gapless Majorana excitations, called the Majorana zero modes~\cite{Kitaev_ANP2006,Willans_PRB2011}.
In the present study, we introduce not static but time-dependent local magnetic fields at one edge and investigate 
how the excited spin polarizations propagate through the system.
From the comprehensive analysis of the spin-component dependence and the comparison with the results for magnetically ordered phases, we 
show that the dynamical spin transport is a good probe for both itinerant Majorana fermions and localized fluxes. 
Our results give 
an insight on the way of 
creating and controlling
of the fractional excitations via the spin degree of freedom. 

The organization of this paper is as follows. 
In Sec.~\ref{sec:model}, we introduce the model and the setup used in this study, 
with the details of real-time evolution and the definitions of 
static and dynamical spin correlations.
In Sec.~\ref{sec:phase}, we show the phase diagram and the static spin correlations in our 
model with edges.
In Sec.~\ref{sec:timedep}, we show how 
an AC local magnetic field at the edge induces
the spin polarization at the opposite edge of the system in
the ferromagnetic phase, the Kitaev QSL, and 
the stripy phase. 
In Sec.~\ref{sec:dynamics},  
we analyze the results in comparison with
the dynamical spin correlations between edges, and discuss the origin of the interedge dynamical spin transport. 
Section~\ref{sec:summary} is devoted to a summary.

\section{Model and Method}
\label{sec:model}

\begin{figure}[t] 
\begin{center} 
\includegraphics[width=0.5 \textwidth]{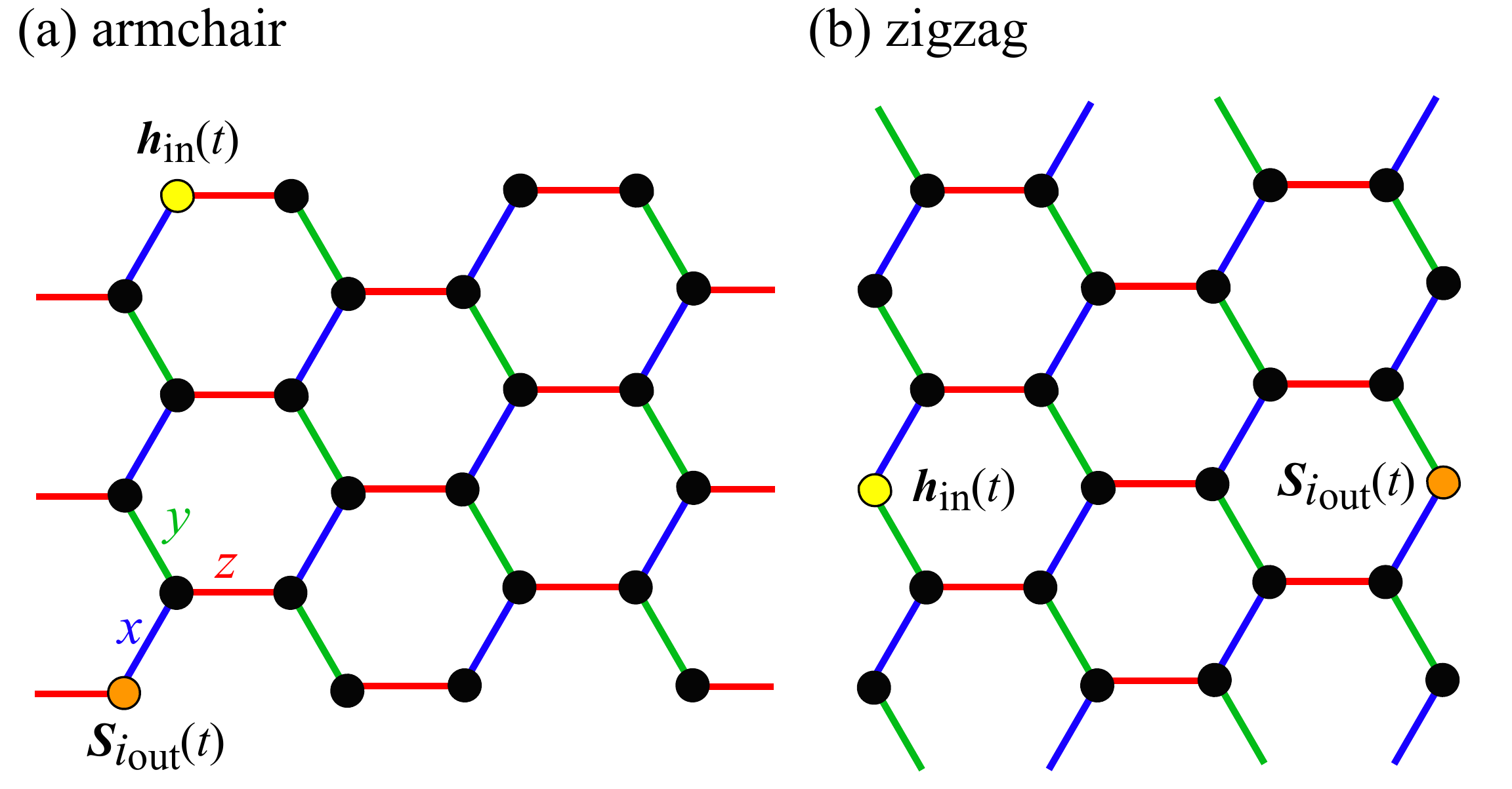}
\caption{Schematic pictures of the Kitaev-Heisenberg model in Eq.~\eqref{eq:Kitaev-Heisenberg} with 
(a)~the armchair edges and (b)~the zigzag edges.
In (a) [(b)], we set the periodic and open (open and periodic) boundary conditions in the horizontal and vertical directions, respectively. 
The blue, green, and red bonds represent the
$x$, $y$, and $z$ bonds for the Kitaev interaction, respectively.
We apply an 
AC magnetic field $\vec{h}_{\rm in}(t)$ 
in the [111] direction in spin space at one site 
on the edge shown by the yellow circles, 
and observe spin polarization at a site on 
the opposite edge shown by the orange circles; see Eqs.~\eqref{eq:H_t}, \eqref{eq:h_in}, and \eqref{eq:spin_pol}. 
}
\label{fig:lattice}
\end{center}
\end{figure}

In this paper, we employ the 
Kitaev-Heisenberg model, whose Hamiltonian is given by 
\begin{align}
\hat{\mathcal{H}}_{\rm KH}=K\sum_{\nu}\sum_{\langle i,j\rangle_\nu}\hat{S}_{i}^{\nu}\hat{S}_{j}^{\nu}
+J\sum_{\langle i,j\rangle}\hat{\vec{S}}_{i}\cdot\hat{\vec{S}}_{j},\label{eq:Kitaev-Heisenberg}
\end{align}
where $\hat{S}_i^\nu$ denotes the $\nu$th component of 
the spin-1/2 operator at {$i$th} site: $\hat{\vec{S}}_i = (\hat{S}_i^x, \hat{S}_i^y, \hat{S}_i^z)$. 
The first term represents the bond-dependent Ising-type 
interaction, called the Kitaev interaction, where 
$\langle i,j\rangle_\nu$ represents the nearest-neighbor $\nu(=x,y,z)$ bonds on the honeycomb lattice, 
and the second term represents the spin-isotropic Heisenberg 
interaction for all the nearest-neighbor bonds; see Fig.~\ref{fig:lattice}. 
Following the previous studies~\cite{Chaloupka_PRL2010,Chaloupka_PRL2013},
we parametrize the two 
coupling constants as 
\begin{align}
\left(K,J\right) = \left(\sin \alpha, \frac12 \cos \alpha \right). 
\end{align}
Note that the amplitudes of interactions are halved 
compared to the previous ones so that $|K|=1$  
in the pure Kitaev cases with $\alpha={\pi/2}$ {and} $3\pi/2$.
In the following, we focus on the 
range of $\pi\leq\alpha\leq7\pi/4$ where the 
Kitaev interaction is 
ferromagnetic. 
In this region, 
the bulk system 
with the periodic boundary conditions 
shows three phases in the ground state~\cite{Chaloupka_PRL2013}: 
the ferromagnetic phase for 
$\pi\leq\alpha\lesssim1.40\pi$, the Kitaev 
QSL phase {for} 
$1.40\pi\lesssim\alpha\lesssim1.58\pi$, and the stripy phase for 
$1.58\pi\lesssim\alpha\lesssim1.81\pi$.

To study spin correlations and dynamics on the edges, we consider the model in Eq.~\eqref{eq:Kitaev-Heisenberg} 
on a strip with the open boundary condition in one direction and the periodic boundary condition in the other. 
There are two types of such strips. 
One has the so-called 
armchair type edges on the open boundaries, and the other has the so-called zigzag edges. 
Figure~\ref{fig:lattice} 
displays these two types for 24-site clusters used in the following calculations.
For both clusters, we examine how 
a time-dependent local magnetic field on one edge
induces spin polarization at the other edge.
Specifically, we apply an AC magnetic field in the [111] direction in spin space at 
$i_{\rm in}$th site on the edge (shown by the yellow circle in Fig.~\ref{fig:lattice}) as
\begin{align}
\hat{\mathcal{H}}(t)=\hat{{\mathcal{H}}}_{\rm KH}+\vec{h}_{\rm in}(t)\cdot\hat{\vec{S}}_{{i_{\rm in}}},
\label{eq:H_t}
\end{align}
with
\begin{align}
\vec{h}_{\rm in}(t)=
h{\bm{e}}_c \cos{\left(\Omega t\right),} 
\label{eq:h_in}
\end{align}
where $h$ is the amplitude of the AC field, 
${\bm{e}}_c=(1,1,1)/\sqrt{3}$, 
and $\Omega=2\pi/T$ is the frequency of the 
AC field ($T$ represents the oscillation period).
We 
take the magnetic 
field along the [111] direction since it coupled to 
all spin components. 
For this Hamiltonian, we solve the 
time-dependent Schr\"{o}dinger equation given by
\begin{align}
i\frac{d\ket{\Phi(t)}}{dt}=\hat{\mathcal{H}}(t)\ket{\Phi(t)}, 
\label{eq:t-dep_Schrodinger}
\end{align}
starting from the initial condition of
$\ket{\Phi(t=0)}=\ket{\Phi_{\rm GS}}$,
where $\ket{\Phi_{\rm GS}}$ is the normalized ground state of $\hat{{\mathcal{H}}}_{\rm KH}$.
The spin polarization 
on the opposite edge is calculated as
\begin{align}
S^{\nu}_{i_{{\rm out}}}(t)=\ev*{\hat{S}^{\nu}_{i_{{\rm out}}}}{\Phi(t)}, 
\label{eq:spin_pol}
\end{align}
where $i_{\rm out}$ denotes the site on the other edge 
directly opposite to the $i_{\rm in}$th site (shown by the orange circle in Fig.~\ref{fig:lattice}). 
In the following calculations, we take $h=0.05$ in Eq.~\eqref{eq:h_in} and 
solve Eq.~\eqref{eq:t-dep_Schrodinger} by using 
$\HPhi$~\cite{Kawamura_CPC2017}; we discretize the time with $\Delta t=0.05$, 
which is small enough to preserve the unitarity of real-time evolution. 

In addition to the real-time dynamics of the spin polarization, 
we calculate the static and dynamical spin correlations 
between the two edges for the ground state $\ket{\Phi_{\rm GS}}$, 
which are defined by
\begin{align}
    {C}_{\rm edge}^{\nu\nu}=\ev*{\hat{S}^{\nu}_{i_{\rm in}}\hat{S}^{\nu}_{i_{\rm out}}}{\Phi_{\rm GS}},
    \label{staticS}
\end{align}
and
\begin{align}
{C}_{\rm edge}^{\nu\nu}(\omega)=\frac{1}{2\pi}\int_{-\infty}^\infty\ev*{\delta\hat{S}^{\nu}_{i_{\rm in}}(t)\delta\hat{S}^{\nu}_{i_{\rm out}}}{\Phi_{\rm GS}}e^{i\omega t} dt,
\label{dynamicalS}
\end{align}
respectively, where 
$\delta\hat{S}^{\nu}_{i}=\hat{S}^{\nu}_{i}-\ev*{\hat{S}^{\nu}_{i}}{\Phi_{\rm GS}}$ {and $\delta\hat{S}^{\nu}_{i_{\rm in}}(t)=e^{i\hat{\mathcal{H}}_{\rm KH}t}\delta\hat{S}^{\nu}_{i_{\rm in}} e^{-i\hat{H}_{\rm KH}t}$}.
Note that in Eq.~\eqref{dynamicalS} we consider correlations 
between the deviations from the expectation values for the ground state to 
subtract the elastic components in the presence of magnetic ordering. 
In the actual calculations of Eq.~\eqref{dynamicalS}, 
we employ the following 
formula in the spectral representation: 
\begin{align}
&{C}_{\rm edge}^{\nu\nu}(\omega)\nonumber\\
&=-\frac{1}{4\pi}{\rm Im}\left[\ev*{(E_{\rm GS}-\hat{\mathcal{H}}_{\rm KH}+\omega+i\eta)^{-1}}{\Psi_+}\right.\notag\\
&\qquad\qquad\quad \left.-\ev*{(E_{\rm GS}-\hat{H}_{\rm KH}+\omega+i\eta)^{-1}}{\Psi_-}\right],
\label{eq:dynamicalS_specrep}
\end{align}
where $\ket{\Psi_\pm}=(\delta\hat{S}^{\nu}_{i_{\rm in}}\pm\delta\hat{S}^{\nu}_{i_{\rm out}})\ket{\Phi_{\rm GS}}$, 
$E_{\rm GS}$ 
is the ground-state energy, and $\eta$ is an infinitesimal positive constant; 
we take $\eta=0.05$ in the following calculations.
We calculate Eq.~\eqref{eq:dynamicalS_specrep} using the continued-fraction expansion based on the Lanczos method. 

In the calculations of Eqs.~\eqref{staticS} and \eqref{eq:dynamicalS_specrep}, 
we apply a weak static magnetic field to all the spins at the edge 
on the $i_{\rm in}$th side with $\bm{h}_s=0.005 {\bm{e}}_c$ to lift the ground-state degeneracy in 
the ferromagnetic Heisenberg model with $\alpha=\pi$ and 
the pure Kitaev model with $\alpha=3\pi/2$ (see Appendix~\ref{ap:degeneracy}). 
For the other cases, the ground state is not degenerate, 
but we apply the same weak field for comparison.

\section{Results}
\subsection{Phase diagram and static spin correlations}
\label{sec:phase}

\begin{figure}[t] 
\begin{center} 
\includegraphics[width=0.5 \textwidth]{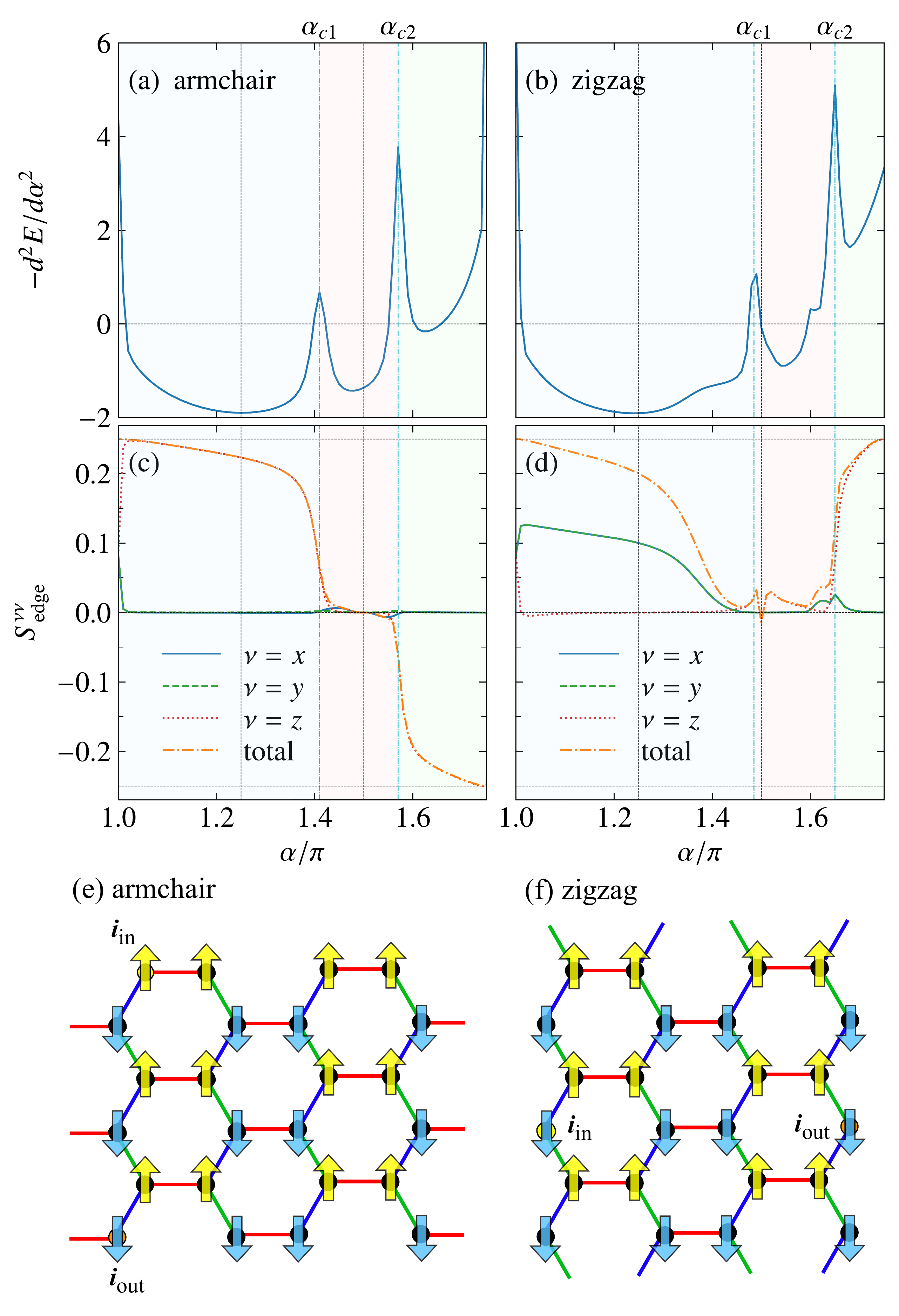}
\caption{$\alpha$ dependences of the 
second derivative of the ground-state energy 
for the systems with (a) the armchair edges and (b) the zigzag edges.
Corresponding 
static interedge spin correlations 
are plotted in (c) and (d).
(e) and (f) show the schematic pictures of the stripy order in the case of armchair and zigzag edges, respectively. 
}
\label{fig:alpha_dep}
\end{center}
\end{figure}

\label{subsubsec:armchair}
Before going into the spin dynamics, we discuss the ground states 
of the clusters with edges shown in Fig.~\ref{fig:lattice}. 
Figures~\ref{fig:alpha_dep}(a) and {\ref{fig:alpha_dep}}(b) show the 
$\alpha$ dependences of the second derivative of the
ground-state energy for {the} systems with the armchair and 
zigzag edges, respectively. 
The two peaks at $\alpha=\alpha_{c1}$ and $\alpha=\alpha_{c2}$ indicate 
phase transitions between the Kitaev 
QSL and the magnetically ordered phases.
Figures~\ref{fig:alpha_dep}(c) and {\ref{fig:alpha_dep}}(d) display 
the static interedge spin correlations defined by Eq.~\eqref{staticS}. 
From these data, we identify three different phases: the ferromagnetic phase with a 
positive spin correlation {for $\alpha<\alpha_{c1}$}, 
the Kitaev QSL with almost zero correlation {for $\alpha_{c1}<\alpha<\alpha_{c2}$}, and the 
stripy phase with a negative (positive) correlation for the system with the armchair (zigzag) edges {for $\alpha>\alpha_{c2}$}. 
The antiferromagnetic and ferromagnetic spin correlations 
in the stripy phase are understood from the schematic pictures 
in Figs.~\ref{fig:alpha_dep}(e) and {\ref{fig:alpha_dep}}(f), respectively.
We note that in the ferromagnetic and stripy phases 
the spin correlations are dominant in a specific spin component 
$S^z$ due to the presence of edges, 
except for $\alpha=\pi$ where the ground state is degenerate 
in the absence of the weak magnetic field $\bm{h}_s$.

Our phase diagrams obtained for the clusters with edges are 
nearly identical to that for the same size cluster under 
the periodic boundary conditions~\cite{Chaloupka_PRL2013}.
For the system with the armchair (zigzag) edges, 
we find that the phase boundary between the ferromagnetic and Kitaev QSL phases is at 
$\alpha_{c1}
\simeq 
1.41\pi$ ($1.49\pi$) and that between the Kitaev QSL and stripy phases is at 
$\alpha_{c2}
\simeq 
1.57\pi$ ($1.65\pi$).
These estimates 
are close to those for the clusters with the periodic boundary conditions, 
$\alpha_{c1} \simeq 1.40\pi$ and $\alpha_{c2} \simeq 1.58\pi$~\cite{Chaloupka_PRL2013}. 
This indicates that the bulk properties are not much affected by the 
introduction of edges even for clusters of this size. 
In the following sections, we will compute the spin dynamics in the three phases: 
For the ferromagnetic, 
Kitaev QSL, and stripy phases, we take $\alpha=1.25\pi$, 
$1.52\pi$, and $1.67\pi$, respectively, for both cases of the armchair and zigzag edges. 

Let us comment on the symmetry of the two clusters in Fig.~\ref{fig:lattice}. 
In the bulk system of the Kitaev-Heisenberg model, there is a four-sublattice transformation 
which does not change the form of the Hamiltonian with 
replacing $K$ and $J$ by $K+J$ and $-J$, respectively~\cite{Chaloupka_PRL2013}.
This transformation leads to the relation between the phase boundaries as $\tan\alpha_{c2}=-\tan\alpha_{c1}-1$.
In the system with the armchair edges, the relation 
holds for our estimates of $\alpha_{c1}$ and $\alpha_{c2}$, since the cluster 
respects the four-sublattice symmetry.
In contrast, in the case of the zigzag edges, 
the symmetry is lost, and $\alpha_{c1}$ and $\alpha_{c2}$ do not satisfy the relation. 

\begin{figure*}[t] 
\begin{center} 
\includegraphics[width=1 \textwidth]{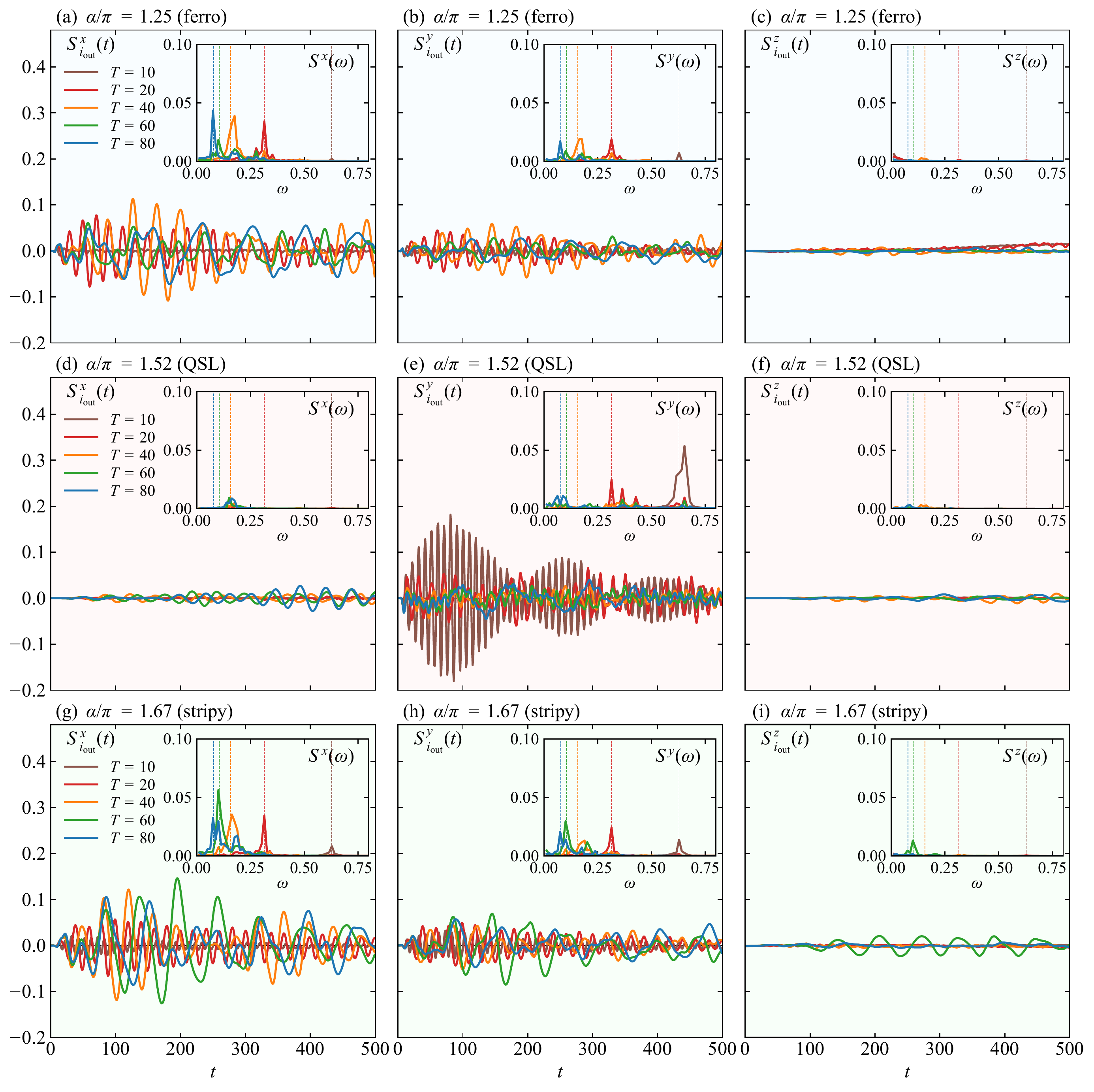}
\caption{Time evolution of the spin polarization 
$S_{{{i_{\rm out}}}}^{\nu}(t)$
in Eq.~\eqref{eq:spin_pol} for the system with 
armchair edges in 
(a)--(c)~the ferromagnetic phase at $\alpha{/\pi}=1.25$, (d)--(f)~{the Kitaev QSL phase at} $\alpha{/\pi}=1.52$, 
and (g)--(i)~the stripy phase at $\alpha{/\pi}=
1.67$: (a),(d),(g) $\nu=x$, (b),(e),(h) $\nu=y$, and (c),(f),(i) $\nu=z$.
The data for the input oscillation periods 
$T=10$, $20$, $40$, $60$, and $80$ are shown. 
The insets display the corresponding Fourier transformed
spin polarizations in Eq.~\eqref{eq:S_omega}. 
The vertical dashed lines denote the frequencies corresponding to the values of $T$, $\omega=\Omega = 2\pi/T$. 
}
\label{fig:arm_all}
\end{center}
\end{figure*}

\subsection{Real-time spin dynamics}
\label{sec:timedep}
We now turn to discuss how the spin at the edge is 
polarized when the AC magnetic field is applied to the spin at the other edge. 
Below, we present the results for the systems with 
armchair and zigzag edges in Secs.~\ref{subsubsec:armchair} and \ref{subsubsec:zigzag}, respectively, 
and discuss the interedge spin resonance in the Kitaev QSL in Sec.~\ref{subsubsec:resonance}. 

\subsubsection{Armchair edge}
\label{subsubsec:armchair}

Figure~\ref{fig:arm_all} displays the time evolution of spin polarization 
at the $i_{\rm out}$th site in Eq.~\eqref{eq:spin_pol} for the system 
with armchair edges in Fig.~\ref{fig:lattice}(a). 
In the main panels, we show the results for the period of the oscillating field, 
$T=10$, $20$, $40$, $60$, and $80$. 
Meanwhile, in the insets, we plot the Fourier transformed spectra obtained by
\begin{align}
S^{\nu}(\omega)=\abs{\frac{2}{t_{\rm max}}\int_{0}^{t_{\rm max}}S_{{{i_{\rm out}}}}^{\nu}(t)e^{{i\omega t}}dt},
\label{eq:S_omega}
\end{align}
where we take $t_{\rm max}=500$ 
so that 
the lowest-energy scale $2\pi/t_{\rm max}$ 
$\sim 0.013$
is well below 
the excitation energy of the localized fluxes ($\sim 0.07$)
in the pure Kitaev model~\cite{Kitaev_ANP2006}.

We first discuss the results in the ferromagnetic 
phase shown in Figs.~\ref{fig:arm_all}(a)--\ref{fig:arm_all}(c). 
In this case, both 
$S_{{{i_{\rm out}}}}^{x}(t)$ and  $S_{{{i_{\rm out}}}}^{y}(t)$ show 
considerable oscillations, while $S_{{{i_{\rm out}}}}^{z}(t)$ does not.
These behaviors are understood from the spin ordering in the ground state: 
As shown in Fig.~\ref{fig:alpha_dep}(c), the spins are 
ferromagnetically ordered in the $z$ direction, 
for which fluctuations 
appear dominantly in the transverse components, 
$S_{{{i_{\rm out}}}}^{x}$ and $S_{{{i_{\rm out}}}}^{y}$, 
rather than the longitudinal one $S_{{{i_{\rm out}}}}^{z}$. 
In the Fourier transformed spectra shown in the insets, 
we find that the dominant $S^{x}(\omega)$ and $S^{y}(\omega)$ 
always show a peak around $\omega=\Omega=2\pi/T$. 
This result indicates that the 
spin polarization is induced dominantly at
the same frequency of the input AC magnetic field.

Next, we turn to the results in the Kitaev QSL phase shown in Figs.~\ref{fig:arm_all}(d)--\ref{fig:arm_all}(f).
In contrast to the above ferromagnetic case, we find that  
only $S_{{{i_{\rm out}}}}^{y}(t)$ shows considerable oscillations,
while the others do not. This behavior can be understood
from the fractional excitations in 
the Kitaev QSL as follows. 
In the exact solution for the pure Kitaev model, 
as mentioned in Sec.~\ref{sec:intro}, the spins are fractionalized into 
itinerant Majorana fermions and localized fluxes. 
The former has gapless excitations, while 
the latter is gapped~\cite{Kitaev_ANP2006}. 
The spin excitation is a composite of these two, and hence gapped. 
Indeed, the spin excitations by $\hat{S}^x$ or $\hat{S}^z$ at the $i_{\rm in}$ or $i_{\rm out}$th site are gapped since these spin operators do not commute with the flux operators defined by products of six spins on the hexagons including the $i_{\rm in}$ or $i_{\rm out}$th site~\cite{Kitaev_ANP2006}. 
This suppresses $S_{i_{\rm out}}^x(t)$ and $S_{i_{\rm out}}^z(t)$ in Figs.~\ref{fig:arm_all}(d) and \ref{fig:arm_all}(f), respectively. 
In contrast, $\hat{S}^y$ at the $i_{\rm in}$ or $i_{\rm out}$th site commutes with the flux operators, since the hexagons lack the $y$ bond. 
In addition to the hexagonal fluxes, 
in the cluster with the armchair edges,
there are additional flux operators defined {only} by the {edge} spins. 
While $\hat{S}^y$ at the $i_{\rm in}$ or $i_{\rm out}$th site do not commute 
with these fluxes, the spin excitations remain gapless because of 
the degeneracy in the ground state (see Appendix~\ref{ap:degeneracy}).  
These allow the excitation by 
$\hat{S}_{i_{\rm in}}^{y}$ to yield 
long-range spin propagation via the gapless itinerant quasiparticles and induce 
$S_{i_{\rm out}}^y(t)$ in Fig.~\ref{fig:arm_all}(e). 
Although the above argument is valid only for the pure Kitaev case,
similar behavior is expected to appear in the Kitaev QSL phase in the presence of weak Heisenberg interactions. 
This is the reason why only $S_{{{i_{\rm out}}}}^{y}(t)$ shows significantly 
large oscillations in Figs.~\ref{fig:arm_all}(d)--\ref{fig:arm_all}(f).

The resonant behaviors in the Kitaev QSL phase exhibit the following characteristics. 
First, while $S^y(\omega)$ always shows a peak  
around the input frequency  
$\Omega$ as in the 
ferromagnetic case,  
the peak height does not decrease but rather increases with $\omega$, 
as shown in the inset of Fig.~\ref{fig:arm_all}(e). 
This characteristic behavior will be discussed in Sec.~\ref{subsubsec:resonance}. 
Second, we note that a weak Heisenberg interaction is crucial for the long-range spin propagation since the ground state degeneracy in the pure Kitaev case with $\alpha/\pi=1.5$ prohibits the propagation, as we will discuss in detail in Sec.~\ref{sec:dynamics}. 
Finally, the above argument also allows 
static interedge spin correlation 
in the $y$ direction,
$S^{yy}_{\rm edge}$, also to develop, but it is almost zero 
as shown in Fig.~\ref{fig:alpha_dep}(c).
This indicates that the interedge spin correlations in the Kitaev QSL 
can only be dynamically enhanced.

Lastly, we discuss the results in the stripy phase
shown in Figs.~\ref{fig:arm_all}(g)--\ref{fig:arm_all}(i).
In this case, the results are similar to the ferromagnetic case in Figs.~\ref{fig:arm_all}(a)--\ref{fig:arm_all}(c). 
The reason is common: As shown in Fig.~\ref{fig:alpha_dep}(c), 
the spins are 
antiferromagnetically ordered in the $z$ direction 
in this stripy phase, and hence, the transverse components $S^x_{i_{\rm out}}(t)$ and $S^y_{i_{\rm out}}(t)$ 
are induced dominantly at the input frequency. 

\subsubsection{Zigzag edge}
\label{subsubsec:zigzag}
\begin{figure*}[t] 
\begin{center} 
\includegraphics[width=1 \textwidth]{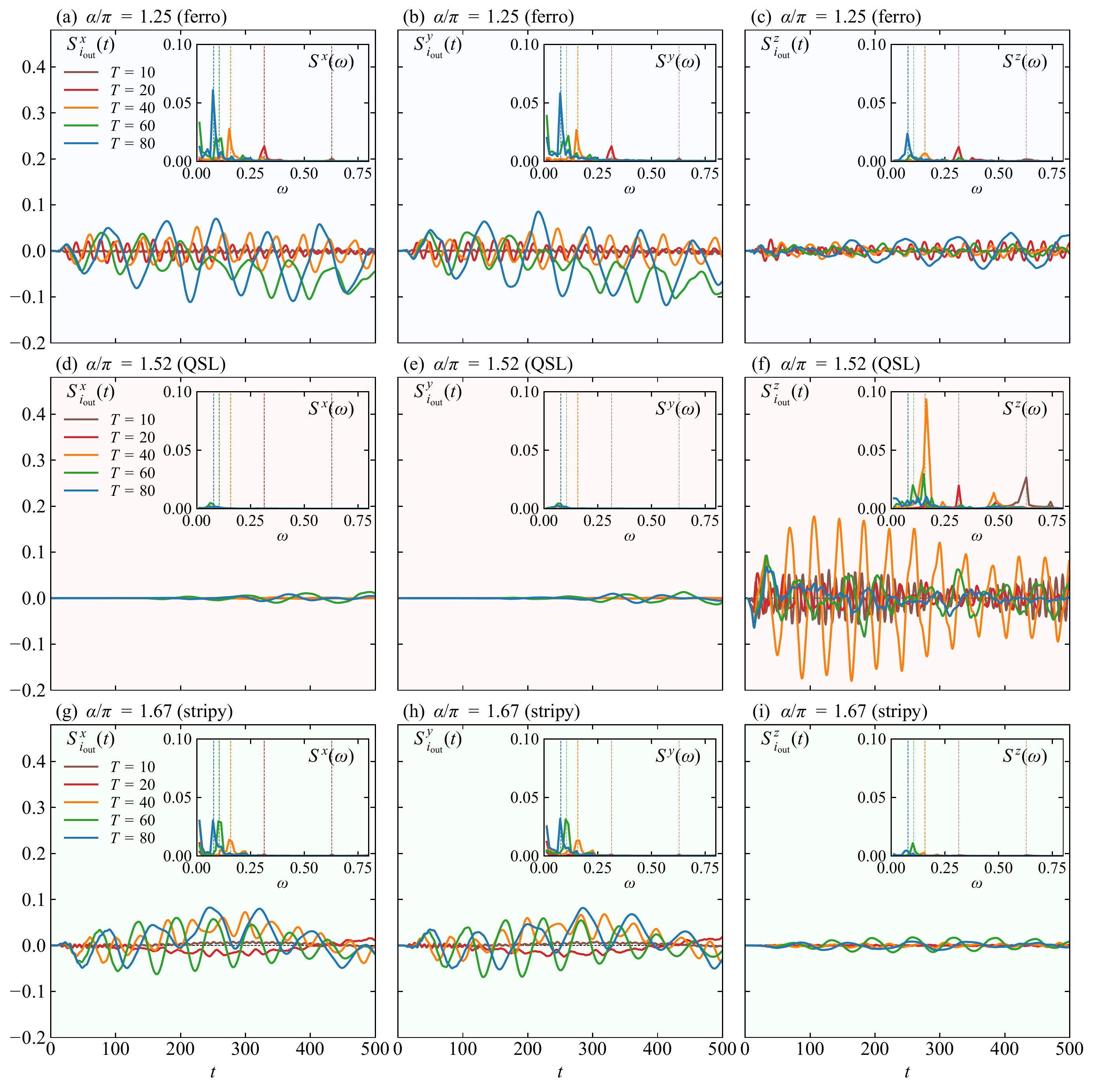}
\caption{Corresponding plots to Fig.~\ref{fig:arm_all} for the system with zigzag edges.}
\label{fig:zig_all}
\end{center}
\end{figure*}

Figure{~\ref{fig:zig_all}} 
displays the results 
for the system with the zigzag edges in Fig.~\ref{fig:lattice}(b), 
obtained by the same conditions for the armchair case.
For the ferromagnetic and stripy phases shown in Figs.~\ref{fig:zig_all}(a)--\ref{fig:zig_all}(c) 
and \ref{fig:zig_all}(g)--\ref{fig:zig_all}(i), respectively, we find similar tendency to the armchair case: 
The spin polarizations in the $x$ and $y$ directions are induced significantly, while that in the $z$ direction is rather suppressed. 
This is again understood from the fact that
the ordered moments in each phase appear in the $z$ direction as shown in Fig.~\ref{fig:alpha_dep}({d}).

Meanwhile, in the Kitaev QSL phase, as 
shown in Figs.~\ref{fig:zig_all}(d)--\ref{fig:zig_all}(f), we find that
$S_{{{i_{\rm out}}}}^{z}(t)$ shows significant large oscillations,
while $S_{{{i_{\rm out}}}}^{x}(t)$ and $S_{{{i_{\rm out}}}}^{y}(t)$ do not. 
This can also be understood from the flux excitations discussed 
above for the armchair case. In the current case, the zigzag edges lack 
the $z$ bonds, and hence, the $S^z$ components can be excited without the gapped 
excitations 
owing to the ground state degeneracy; see Appendix~\ref{ap:degeneracy}. 
Interestingly, the amplitude of $S_{{{i_{\rm out}}}}^{z}(t)$  
varies nonmonotonically with $T$ and
takes the maximum value around $T=40$, as shown in 
Fig.~{\ref{fig:zig_all}(f)}. 
In this case also, we observe the peaks in 
$S^z(\omega)$ at almost the input frequencies, 
as shown in the inset of Fig.~\ref{fig:zig_all}(f), 
while the peak heights show nonmonotonic $\omega$ dependence.
These features will be discussed in the next section.

\subsubsection{Interedge resonance}
\label{subsubsec:resonance}

\begin{figure*}[t] 
\begin{center} 
\includegraphics[width=1 \textwidth]{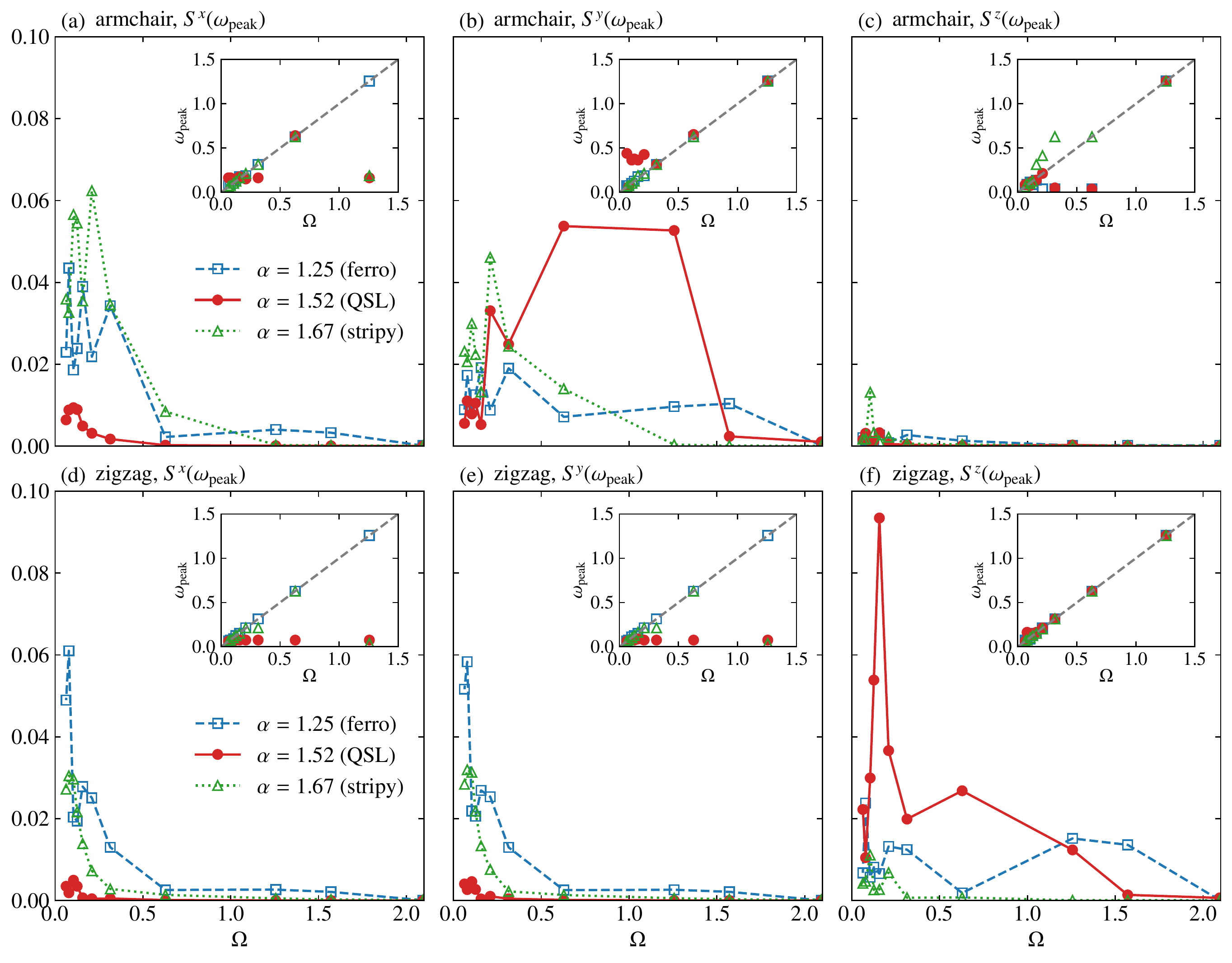}
\caption{Maximum values of the Fourier transformed spin polarizations
$S^{\nu}(\omega_{\rm peak})$ 
for the input oscillating magnetic field with $\Omega = 2\pi/T$ 
in the systems with (a){--}(c) the armchair edges and 
(d){--}(f) the zigzag edges: (a) and (d) $\nu=x$, (b) and (e) $\nu=y$, and (c) and (f) $\nu=z$.
The insets display $\Omega$ dependences of $\omega_{\rm peak}$.
The gray dashed line shows the relation $\omega_{\rm peak}=\Omega$. 
}
\label{fig:peak}
\end{center}
\end{figure*}

In Fig.~\ref{fig:peak}, we show 
$\Omega=2\pi/T$  
dependences of the
maximum values of $S^{\nu}(\omega)$ ($\nu=x,y,z$) {within the range of $2/t_{\rm max}<\omega/2\pi\leq0.5$}, 
which are represented as $S^{\nu}(\omega_{\rm peak})$,  
in the systems with (a)-(c) the armchair edges and (d)-(f) the zigzag edges.
We also plot the values of $\omega_{\rm peak}$ as functions of $\Omega$ in each inset. 
Here, we choose the lower limit as twice $2\pi/t_{\rm max}$ to avoid an
artifact near $2\pi/t_{\rm max}$ and the 
upper limit to be sufficiently larger than 
the bandwidth of the dynamical spin correlation~\cite{Knolle2014dynamics}.

In both armchair and zigzag cases, $S^x(\omega_{\rm peak})$ and $S^y(\omega_{\rm peak})$ 
have large values in the ferromagnetic and stripy phases, 
while $S^z(\omega_{\rm peak})$ are suppressed, 
since these phases show the spin orderings along the $z$ direction as mentioned above. 
The values of $S^x(\omega_{\rm peak})$ and $S^y(\omega_{\rm peak})$ decrease 
as $\Omega$ increases. 
In addition, we find that the values of $\omega_{\rm peak}$ are close to $\Omega$ as shown in each inset. 
These are indications of conventional magnetic resonances; 
the spin polarization is induced dominantly at the input frequency, 
as long as the magnetic excitations are available in the frequency range. 

In contrast, the interedge spin resonance behaves differently in the Kitaev QSL phase. 
First of all, we find that the dominant polarizations, $S^y(\omega_{\rm peak})$ 
for the armchair case and $S^z(\omega_{\rm peak})$ for the zigzag case, 
show broad peaks in the wide frequency range up to $\Omega \simeq 1.5$, while the latter shows a sharp 
peak at a smaller $\Omega \simeq 0.2$ as well. 
For both cases, the relation $\omega_{\rm peak} \sim \Omega$ holds, 
except for small $\Omega$ ($\Omega \lesssim 0.2$ for the armchair case and $\Omega \lesssim 0.1$ for the zigzag case); 
the deviation might be due to the finite-size effect. 
The broad responses with $\omega_{\rm peak} \sim \Omega$, 
as well as 
the sharp peak in the zigzag case, can be ascribed to 
the itinerant Majorana fermions, 
whose density of states shows a continuum up to $\omega=1.5$~\cite{Kitaev_ANP2006}. 
This point will be further discussed in Sec.~\ref{sec:dynamics}.

In addition, for the other spin components with suppressed polarizations, we find that $\omega_{\rm peak}$ is 
almost constant irrespective of $\Omega$~\footnote{The data for $S^z(\omega_{\rm peak})$ 
at $\Omega =2\pi/10 \simeq 0.63$ and $S^z(\omega_{\rm peak})$ at $\Omega = 2\pi/5 \simeq 1.26$ 
in the armchair case deviate from the constant behavior, and rather close to $\Omega$. 
In these cases, we also observe peaks around the constant values, 
but the peak heights are slightly smaller than those around $\Omega$.}. 
This behavior could be explained by the flux gap that governs the low-energy excitations in these spin components.
We note that the constant values of $\omega_{\rm peak} \sim 0.2$ 
in the armchair case is larger than the energy of the low-energy
coherent peak at $\omega \sim 0.1$ in the dynamical spin 
structure factor of the pure Kitaev model in the 
thermodynamic limit~\cite{Knolle2014dynamics}, but this might 
also be due to the finite-size effect. 

In the Kitaev QSL, $S^x(\omega_{\rm peak})$ is larger than $S^z(\omega_{\rm peak})$ in the armchair case, 
while $S^x(\omega_{\rm peak})$ and $S^y(\omega_{\rm peak})$ are 
almost the same in the zigzag case. 
This is understood from the geometry of the Kitaev bonds in each cluster. 
In the armchair case, as shown in Fig.~\ref{fig:lattice}(a), 
the $z$ bonds on which $\hat{S}^z$  
components interact via the Kitaev interaction are along the edges 
and perpendicular to the direction from  $i_{\rm in}$ to $i_{\rm out}$, 
which may suppress the interedge spin transport of the $z$ component. 
Meanwhile, in the zigzag case shown in Fig.~\ref{fig:lattice}(b), 
both $x$ and $y$ bonds are along the edges and related with each other by symmetry, 
leading to almost the same interedge resonances assisted by the $z$ bonds connecting them. 

\subsection{Comparison with dynamical spin correlations}
\label{sec:dynamics}

\begin{figure}[t] 
\begin{center} 
\includegraphics[width=0.5 \textwidth]{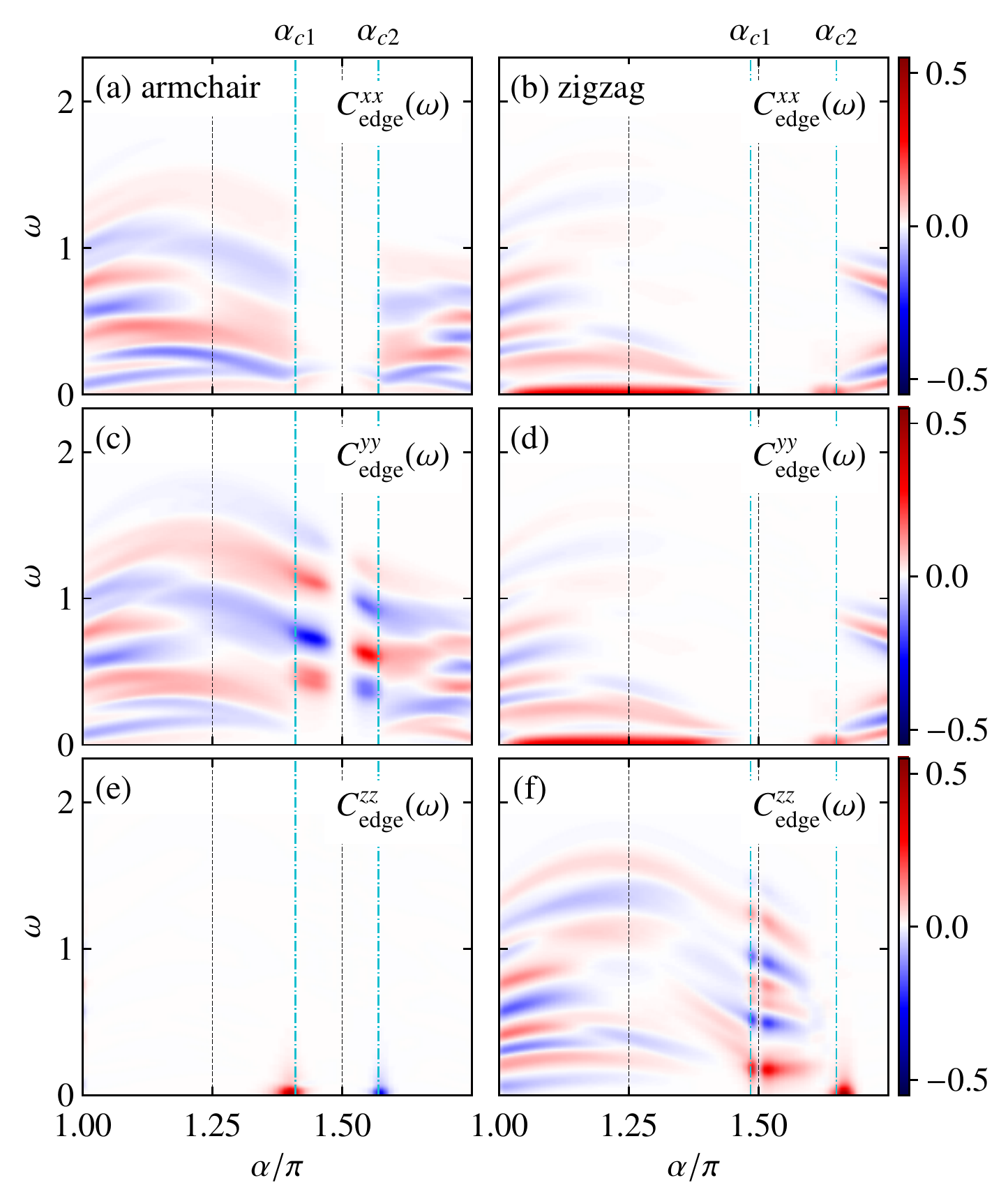}
\caption{Dynamical interedge spin correlations ${C}_{\rm edge}^{\nu\nu}(\omega)$
[Eq.~(\ref{dynamicalS})] 
for (a),(c),(e) the armchair edge and (b),(d),(f) the zigzag edge: 
(a) and (b) $\nu=x$, (c) and (d) $\nu=y$, and (e) and (f) $\nu=z$.
}
\label{fig:dym}
\end{center}
\end{figure}

Let us discuss the characteristic interedge spin resonances in comparison with 
the dynamical interedge spin correlations
${C}_{\rm edge}^{\nu\nu}(\omega)$ defined in Eq.~(\ref{dynamicalS}). 
Figure~\ref{fig:dym} displays ${C}_{\rm edge}^{\nu\nu}(\omega)$
for the armchair and zigzag cases while varying $\alpha$. 
In the following, we show that ${C}_{\rm edge}^{\nu\nu}(\omega)$
explains well the intensities and $\Omega$ dependences of the induced spin polarizations 
in Fig.~\ref{fig:peak}.

In the system with armchair edges, ${C}_{\rm edge}^{xx}(\omega)$ and 
${C}_{\rm edge}^{yy}(\omega)$ show considerable intensities over the broad $\omega$ range, 
whereas ${C}_{\rm edge}^{zz}(\omega)$ is almost zero except for 
the low-$\omega$ weights near the phase boundaries at $\alpha=\alpha_{c1}$ and $\alpha_{c2}$. 
In the ferromagnetic phase 
for $\alpha\leq\alpha_{c1}$ and the stripy phase 
for $\alpha\geq\alpha_{c2}$, this is again consistent with {the fact} that the spin moments are ordered along the $z$ direction. 
A striking difference between 
${C}_{\rm edge}^{xx}(\omega)$ and ${C}_{\rm edge}^{yy}(\omega)$ appears 
in the {Kitaev} QSL phase for $\alpha_{c1}\leq\alpha\leq\alpha_{c2}$;
${C}_{\rm edge}^{yy}(\omega)$ has large 
spectral weights over the broad $\omega$ range,
while ${C}_{\rm edge}^{xx}(\omega)$ 
is almost zero.
Notably, the intensity 
of ${C}_{\rm edge}^{yy}(\omega)$ is stronger 
than those in the ferromagnetic and stripy phases, while it vanishes for the 
pure Kitaev case at $\alpha/\pi=1.5$ because of the degeneracy 
in the ground state (see Appendix~{\ref{sec:DynamicalMajorana}}).
This {strong ${C}_{\rm edge}^{yy}(\omega)$} explains well the broad response in $S^y(\omega_{\rm peak})$ found in Fig.~\ref{fig:peak}(b). 
In addition, we note that ${C}_{\rm edge}^{xx}(\omega)$ has weak intensities at low $\omega \sim 0.1$, as shown in Fig.~\ref{fig:dym}(a). 
This also explains well the small peak in $S^x(\omega_{\rm peak})$ found in Fig.~\ref{fig:peak}(a).

Meanwhile, in the system with zigzag edges, ${C}_{\rm edge}^{\nu\nu}(\omega)$ in the ferromagnetic 
and stripy phases behave qualitatively similarly to those in the armchair case. 
In the Kitaev QSL phase, however, strong intensity appears 
in ${C}_{\rm edge}^{zz}(\omega)$ over the broad $\omega$ range, 
while ${C}_{\rm edge}^{xx}(\omega)$ and ${C}_{\rm edge}^{yy}(\omega)$ are almost absent. 
Again, this explains well the broad response in $S^z(\omega_{\rm peak})$ found in Fig.~\ref{fig:peak}(f). 
Furthermore, the sharp peak at $\omega \sim 0.2$ in Fig.~\ref{fig:peak}(f) is 
also consistent with the strong intensity of ${C}_{\rm edge}^{zz}(\omega)$ in Fig.~\ref{fig:dym}(f).

\begin{figure}[t] 
\begin{center} 
\includegraphics[width=0.5 \textwidth]{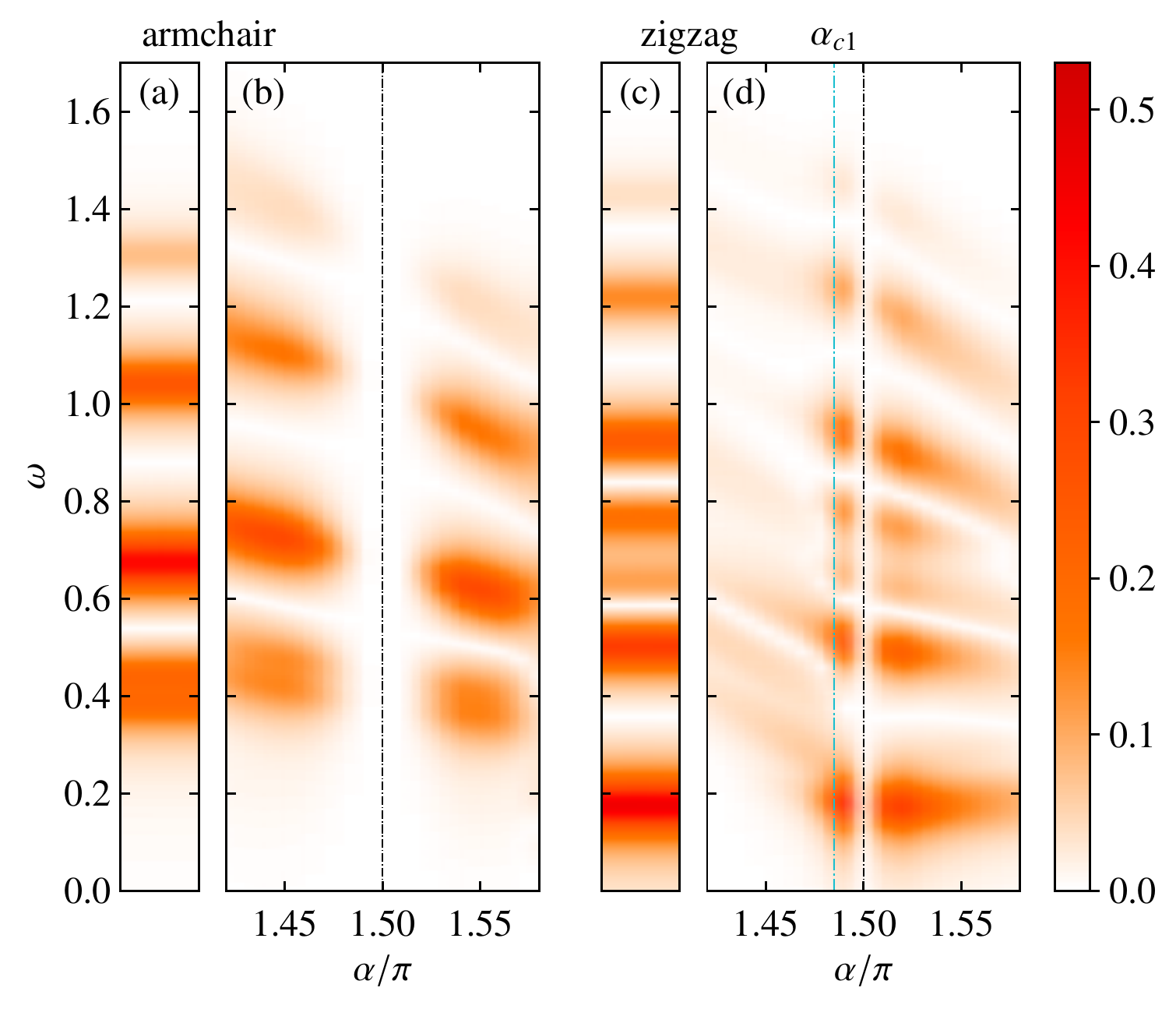}
\caption{Comparison between (a) 
$|{C}_{\rm edge}^{\rm Maj}(\omega)|$ 
calculated by
Eq.~\eqref{eq:MajoranaG} 
{for} the {pure} Kitaev model {at $\alpha/\pi=1.5$} and (b) an enlarged plot of 
$|{C}_{\rm edge}^{yy}(\omega)|$ around $\alpha/\pi=1.5$ 
for the system with the armchair edges.
(c) and (d) The corresponding plots for the zigzag case, 
where $|{C}_{\rm edge}^{zz}(\omega)|$ is plotted in ({d}).
}
\label{fig:hikaku}
\end{center}
\end{figure}

The interedge resonances in the broad $\omega$ range in the Kitaev QSL phase 
are mediated by the itinerant Majorana 
fermions whose excitation spectrum has a continuum in the broad energy range. 
This is explicitly shown by calculating the 
dynamical spin correlations for the pure Kitaev model at $\alpha/\pi = 1.5$ 
by using the Majorana representation, which we denote ${C}_{\rm edge}^{\rm Maj}(\omega)$; 
see Appendix~\ref{sec:DynamicalMajorana} for the details of the calculations. 
Figure~\ref{fig:hikaku} shows the results {of ${C}_{\rm edge}^{\rm Maj}(\omega)$} 
in comparison with ${C}_{\rm edge}^{yy}(\omega)$ and ${C}_{\rm edge}^{zz}(\omega)$ {around $\alpha/\pi=1.5$.} 
Note that {here} we compare their absolute values since the sign of 
{${C}_{\rm edge}^{\rm Maj}(\omega)$ in Eq.~\eqref{eq:MajoranaG}} is not well defined. 
We find that the broad responses of ${C}_{\rm edge}^{yy}(\omega)$ and ${C}_{\rm edge}^{zz}(\omega)$ 
in the vicinity of $\alpha/\pi = 1.5$ appear in the same energy range of 
{${C}_{\rm edge}^{\rm Maj}(\omega)$} with showing similar $\omega$ dependences. 
This indicates that the broad responses in the 
Kitaev QSL phase are dominated by the itinerant Majorana excitations.

\begin{figure}[t] 
    \begin{center} 
    \includegraphics[width=0.5 \textwidth]{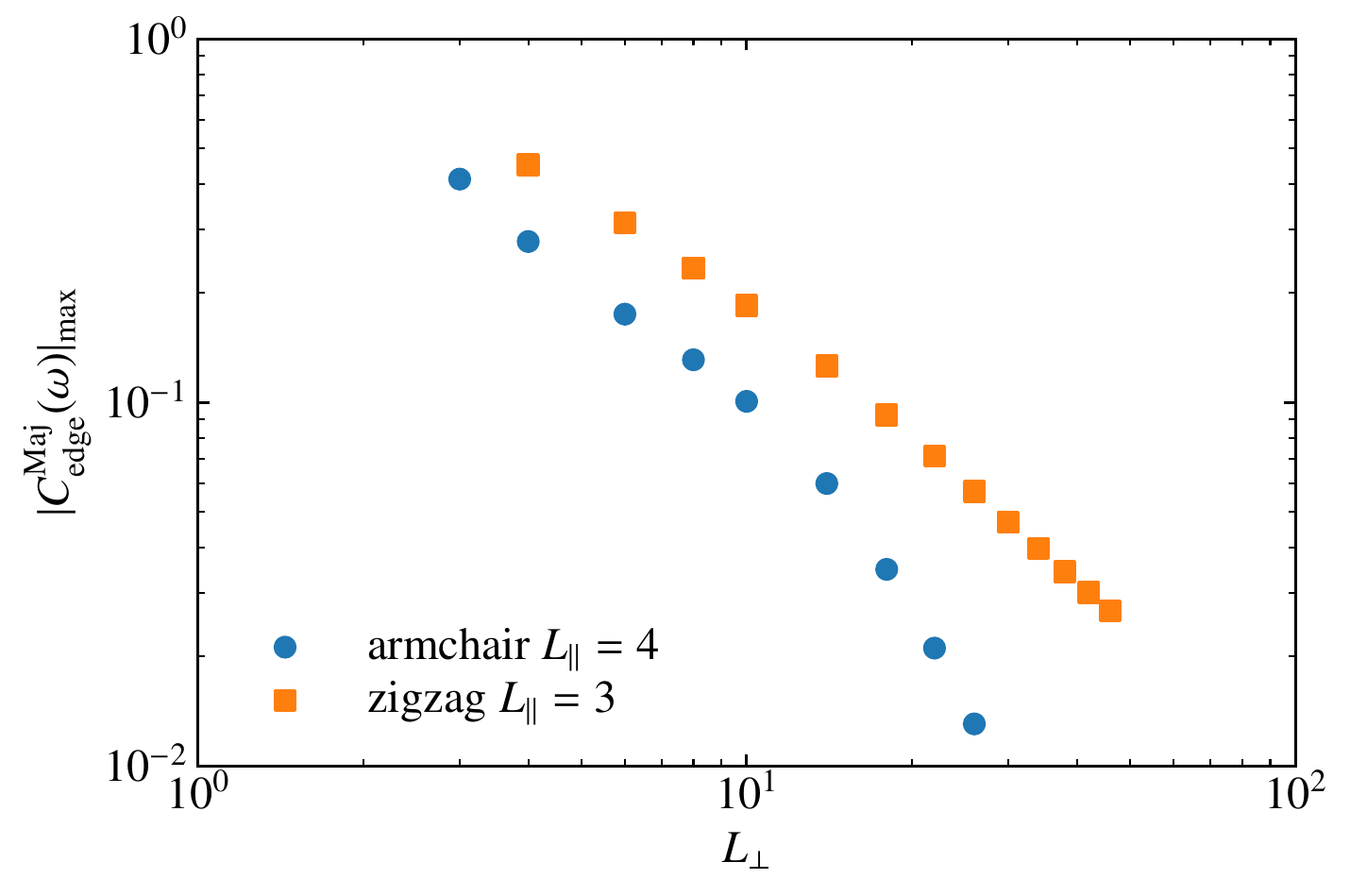}
    \caption{
Maximum of $|{C}_{\rm edge}^{\rm Maj}(\omega)|$ 
as a function of the number of the unit cells in the direction perpendicular to the edges, $L_{\perp}$.
The data are calculated for the clusters with the numbers of the unit cell along the edge, $L_{\parallel}=4$ 
and $3$, for the armchair 
and zigzag cases, respectively; the 
result for the smallest $L_\perp$ in each case corresponds to 
that in Figs.~\ref{fig:hikaku}(a) and \ref{fig:hikaku}(c). 
    }
    \label{fig:Ldep}
    \end{center}
\end{figure}

While our results are limited to the small clusters, we expect that the interedge resonances appear also in larger systems 
since the itinerant Majorana fermions propagate 
over long distances in the Kitaev QSL. 
This is demonstrated by calculating $|{C}_{\rm edge}^{\rm Maj}(\omega)|$ 
while changing the system width. 
Figure~\ref{fig:Ldep} shows the maximum intensity of $|{C}_{\rm edge}^{\rm Maj}(\omega)|$ as a function of 
the number of the unit cells in the direction perpendicular to the edges, $L_\perp$.
We find that the dynamical correlations decay slowly: the zigzag case roughly obeys $\propto 1/L_\perp$, while the armchair case shows crossover from $\propto 1/L_\perp$ to $\propto 1/L_\perp^3$. 
The results appear to be consistent with the Majorana-mediated spin correlations~\cite{Koga_PRB2021,Minakawa_PRL2020}. 
Thus, we believe that, when dominated by the itinerant Majorana fermions, the interedge dynamical spin correlations become long-range in real space, 
even in the presence of weak Heisenberg interactions~\cite{Tikhonov_PRL2011,Mandal_PRB2011,Song_PRL2016}. 

Combining these results with the almost constant behaviors of $\omega_{\rm peak}$ 
irrespective of $\Omega$ for the other suppressed components in Fig.~\ref{fig:peak}, 
we conclude that the interedge spin resonances 
in the Kitaev QSL are good probes of two types of fractional excitations, 
itinerant Majorana fermions and localized fluxes. 
The resonance in the spin component which does not excite the fluxes on hexagons 
leads to broad responses with $\omega_{\rm peak} \sim \Omega$, 
as found in Figs.~\ref{fig:peak}(b) and \ref{fig:peak}(f). 
This is a clear indication of the itinerant Majorana excitations.
Meanwhile, the responses in the other spin components appear around a small constant $\omega_{\rm peak}$. 
This is an indication of the gapped flux excitations. 
We emphasize that weak Heisenberg interactions are essential for the interedge spin resonances 
since all ${C}_{\rm edge}^{\nu\nu}(\omega)$ vanish for the pure Kitaev model 
because of the ground-state degeneracy (see Appendix~\ref{sec:DynamicalMajorana}).

\section{Summary}
\label{sec:summary}
In summary, we have studied how an AC 
local magnetic field at an edge of the system induces 
spin polarizations at the opposite edge
in the Kitaev-Heisenberg model with ferromagnetic Kitaev interactions by using the exact diagonalization. 
We found that in the Kitaev QSL phase 
the spin polarizations are resonantly induced in a particular spin component, in stark contrast to the magnetically ordered phases where conventional magnetic resonances appear in the transverse spin components. 
The spin resonance in the Kitaev QSL shows the following peculiar features, stemming from the fractionalization of spin degree of freedom into two types of fractional excitations, itinerant Majorana fermions and localized fluxes: 
(i) It appears dominantly in the spin component which does not excite flux excitations, 
(ii) the dominant resonance appears in a broad range of frequency, reflecting the continuum of Majorana excitations, 
(iii) it is accompanied by subdominant resonances in the other spin components at a small constant frequency corresponding to the flux excitation gap, 
(iv) both resonances vanish in the exact Kitaev QSL because of the ground-state degeneracy and require weak Heisenberg interactions, and 
(v) they are induced only dynamically, despite the disappearance of
the static spin correlations. 
These results elucidate that the nonlocal spin dynamics in the wide frequency range contains information on both two types of fractional excitations in the Kitaev QSL, which cannot be captured by the spin transport in the 
low-energy limit 
in the previous studies~\cite{Minakawa_PRL2020,Taguchi_JPS2023}.
While our calculations were done for small size clusters, 
the interedge resonance is expected to be observed in larger systems, 
since it is mediated by itinerant Majorana excitations that propagate 
over long distances. 
These results indicate that the interedge dynamical spin resonance 
is useful for probing the two types of 
fractional excitations 
in the Kitaev QSL,
which are usually difficult to observe only from static physical quantities. 

A straightforward experiment would be implemented 
by using a scanning tunneling microscope (STM) tip 
with magnetic atoms or the atomic force microscopy (AFM) to apply an
AC  
magnetic field at the edge and
measure the spin polarization at the opposite edge.
This could be performed, for example, for a thin flake of a 
candidate material $\alpha$-RuCl$_{3}$.  
Similar experiments would be possible in interface or heterostructure of a Kitaev magnet and a ferromagnetic material, where the AC magnetic field can be applied to the edge spins by the ferromagnetic resonance.
Careful analysis of the dynamics in each spin component and its dependence on the edge structure 
would pave the way for creating and controlling the fractional excitations 
in the Kitaev QSL through the spin degree of freedom. 

\begin{acknowledgments}
We wish to thank Y. Kato, K. Fukui, and T. Okubo 
for fruitful discussions.
This work was also supported by the
National Natural Science Foundation of China (Grant No.~12150610462).
TM was supported by Building of Consortia for 
the Development of Human Resources in Science and Technology, MEXT, Japan.
This work was supported by  Grant-in-Aid for Scientific Research
Nos.~19H05825, JP19K03742, and 20H00122 
from the Ministry of Education, Culture, Sports, Science and Technology, Japan.  
It is also supported by JST CREST Grant No.~JPMJCR18T2 
and JST PRESTO Grant No.~JPMJPR19L5.

\end{acknowledgments}

\appendix
\section{Degeneracy in the pure Kitaev model with edges}
\label{ap:degeneracy}

\begin{figure}[t]
\begin{center}
\includegraphics[width=0.5 \textwidth]{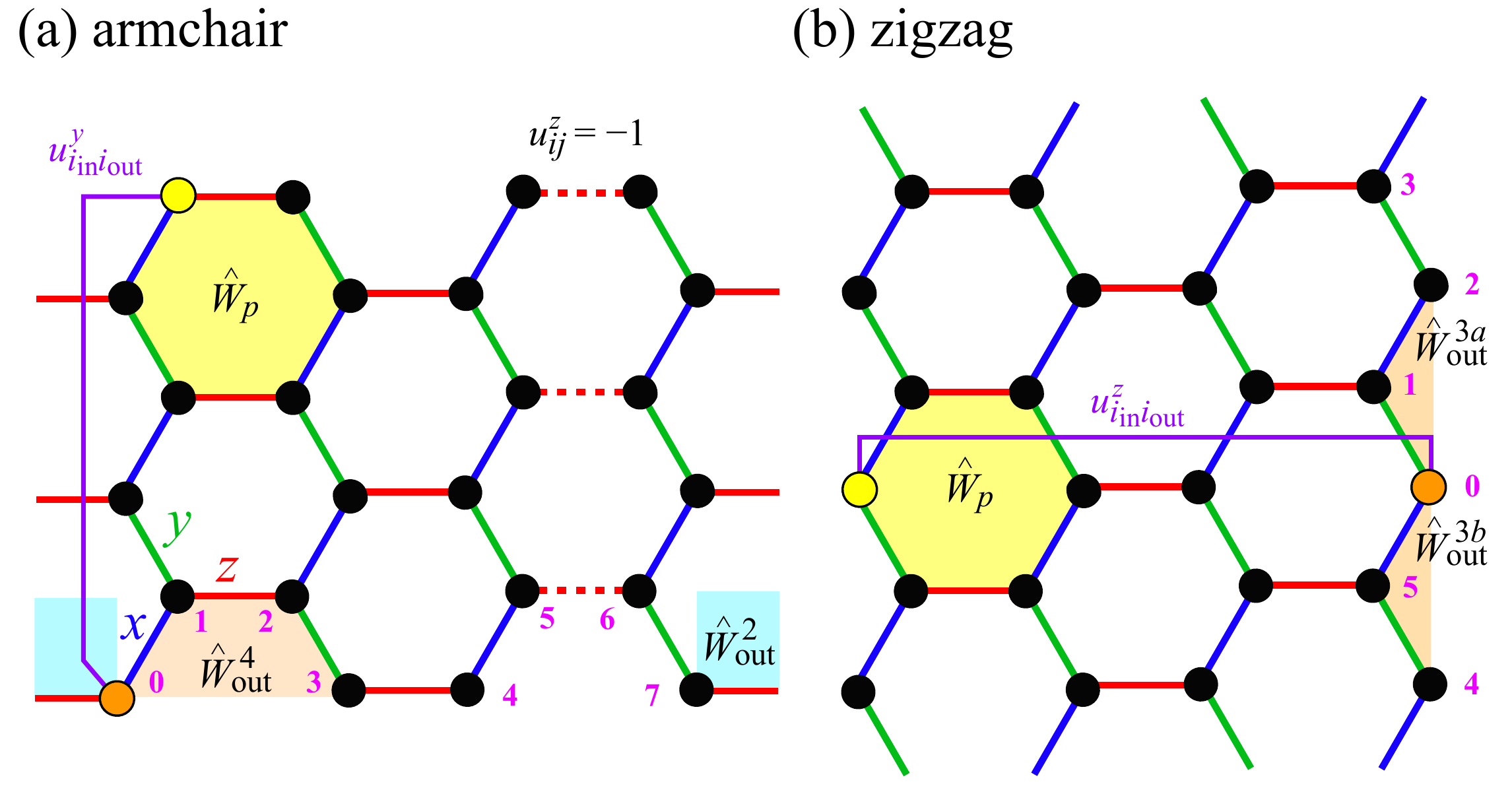}
\caption{
Schematic pictures of the flux operators 
for (a) the armchair edges and (b)~the zigzag edges.
The numbers denote the sites used for the definitions of the fluxes 
in Eq.~\eqref{eq:flux1} and Eq.~\eqref{eq:flux2}.
We show examples of the six-spin flux operator $\hat{W}_{{p}}$,
four-spin ($\hat{W}^{4}$) and two-spin flux operators  ($\hat{W}^{2}$) 
and the three-spin flux operators ($\hat{W}^{3{a}}$ and $\hat{W}^{3b}$). 
We also show the interedge correlations of 
the localized Majorana particles ($u_{i_{\rm in}i_{\rm out}}$)
by the purple lines.
In (a), we represent the $z$ bonds 
where $u_{ij}$ takes -1 with red dashed lines. 
See Appendix \ref{sec:DynamicalMajorana} {for $u_{i_{\rm in}i_{\rm out}}$ and $u_{ij}$}.
}
\label{fig:app_lattice}
\end{center}
\end{figure}

In this Appendix, we show that the ground state of the pure Kitaev model with $\alpha=3\pi/2$ has the degeneracy in both cases of the armchair and the zigzag edges. 
In the pure Kitaev model, one can define the flux operator $\hat{W}_p$ by a product of six spins for each hexagon $p$, which commutes with the Hamiltonian~\cite{Kitaev_ANP2006}. 
We show the examples in Fig.~\ref{fig:app_lattice}; 
note that $\hat{W}_p$ in (a) [(b)] commutes with $\hat{S}_{i_{\rm in}}^y$ ($\hat{S}_{i_{\rm in}}^z$), since the hexagon lacks the $y$ ($z$) bond at the $i_{\rm in}$th site, as discussed in Sec.~\ref{subsubsec:armchair} (\ref{subsubsec:zigzag}). 
In addition {to the six-spin flux operators}, 
{at the} edges {of the system} there are additional flux operators defined only by the edge spins. 
For instance, the flux operators including the output site are given by
\begin{align}
\hat{W}^{4}_{\rm out} 
= 2^{4}\hat{S}^{z}_{0}\hat{S}^{y}_{1}\hat{S}^{x}_{2}\hat{S}^{z}_{3}, 
\quad
\hat{W}^{2}_{\rm out} 
= 2^{2}\hat{S}^{x}_{0}\hat{S}^{y}_{7}, 
\label{eq:flux1}
\end{align} 
for the armchair case, and 
\begin{align}
\hat{W}^{3a}_{\rm out} 
= 2^{3}\hat{S}^{x}_{0}\hat{S}^{z}_{1}\hat{S}^{y}_{2},
\quad
\hat{W}^{3b}_{\rm out} 
= 2^{3}\hat{S}^{y}_{0}\hat{S}^{z}_{5}\hat{S}^{x}_{4}, 
\label{eq:flux2}
\end{align}
for the zigzag case; see Fig.~\ref{fig:app_lattice}. 
Since these flux operators $\hat{W}^q$ ($q=4$, $2$, $3a$, and $3b$) commute 
with the Hamiltonian at $\alpha=3\pi/2$,
the ground state 
$\ket{\Phi_{\rm GS}}$ 
is the eigenstate of 
the flux operators as 
\begin{align}
\hat{W}^{q}_{\rm out}\ket{\Phi_{
{{\rm GS}}}}=
{W} \ket{\Phi_{
{{\rm GS}}}}, 
\label{eq:Wq_eigen}
\end{align}
{where the eigenvalue $W$ takes $+1$ or $-1$.} 
Meanwhile, 
all the eigenstates of the Hamiltonian can be taken to be real 
since the Kitaev Hamiltonian 
does not include the complex elements 
in the conventional 
basis set composed of the eigenstates of 
$\hat{S}_i^z$. 
Therefore, if 
the ground state $\ket{\Phi_{\rm GS}}$ is unique, we obtain
$\ev{\hat{W}^{q}_{\rm out}}{\Phi_{{{\rm GS}}}}=0$
since $\hat{W}^{q}_{\rm out}$ 
is the pure 
imaginary operator including 
a single $\hat{S}^{y}_{i}$. 
This contradicts 
with Eq.~\eqref{eq:Wq_eigen}, meaning that 
the assumption of a unique ground state is incorrect. 
Hence, the ground state of the pure Kitaev model with the armchair and the zigzag
edges must be degenerate. 
For the 24-site clusters 
shown in 
Fig.~\ref{fig:lattice}, we numerically confirm that
the ground state has 
eightfold (fourfold) degeneracy
for the clusters with the armchair (zigzag) edges.
We note that the numbers of the degenerate states can be accounted 
for by the numbers of independent flux-type operators 
traversing the system from one edge to the other and those consisting of edge spins.

\section{Dynamical spin correlations in the pure Kitaev model} 
\label{sec:DynamicalMajorana}
In this Appendix, we 
describe the method to calculate the
dynamical spin correlations  
for the pure Kitaev model in Figs.~\ref{fig:hikaku}(a) and \ref{fig:hikaku}(c). 
We adopt 
the Majorana representation of the Hamiltonian in Eq.~\eqref{eq:Kitaev-Heisenberg} at $\alpha=3\pi/2$, which 
is given by~\cite{Kitaev_ANP2006} 
\begin{align}
    \hat{\mathcal{H}}_{\rm K}=\frac{1}{4}\sum_{\langle i,j\rangle_{\nu}}{u}_{ij}^{\nu} i\hat{c}_i\hat{c}_j,
\end{align}
where the spin operator is represented as $\hat{S}_{i}^{\nu}=\frac{i}{2}\hat{b}_i^\nu \hat{c}_i$ 
by introducing four Majorana fermions $\{\hat{c}_i,\hat{b}_i^x,\hat{b}_i^y,\hat{b}_i^z\}$. 
Here, $u_{ij}^{\nu}={\ev*{\hat{u}_{ij}^{\nu}}{\Phi_{\rm GS}}}
={\ev*{i\hat{b}_i^\nu \hat{b}_j^\nu}{\Phi_{\rm GS}}}${;} 
$\hat{u}_{ij}^{\nu}$  
commutes with the Hamiltonian and
$u_{ij}^{\nu}$ takes $\pm 1$. 

In this Majorana representation, the interedge dynamical spin correlation is given by
\begin{align}
&\ev*{\hat{S}_{i_{\rm in}}^{\nu}(t) \hat{S}_{i_{\rm out}}^{\nu}}{\Phi_{\rm GS}}\notag\\
&\qquad  =-\frac{1}{4}u_{{i_{\rm in} i_{\rm out}}}^{\nu}
{\ev*{ i\hat{c}_{i_{\rm in}}(t)\hat{c}_{i_{\rm out}}}{\Phi_{\rm GS}}},
\label{eq:SS_Majorana}
\end{align}
where $\nu=y$ and $z$ for the armchair and zigzag case, respectively, and 
$u_{{i_{\rm in} i_{\rm out}}}^{\nu}$ 
is defined for the unpaired $\hat{b}_{i_{\rm in}}^\nu$ and $\hat{b}_{i_{\rm out}}^\nu$ on the edges as 
$u_{i_{\rm in} i_{\rm out}}^\nu = \ev*{i\hat{b}_{i_{\rm in}}^\nu \hat{b}_{i_{\rm out}}^\nu}{\Phi_{\rm GS}}$. 
The correlations for the other spin components vanish. 
By substituting Eq.~\eqref{eq:SS_Majorana} to Eq.~\eqref{dynamicalS}, we 
obtain the dynamical spin correlation between the edges in the Majorana representation as 
\begin{align}
{{C}_{\rm edge}^{\rm Maj}}(\omega)
={-}{\frac{{u_{i_{\rm in} i_{\rm out}}^\nu}}{8\pi}}\int_{-\infty}^\infty 
{\ev*{ i\hat{c}_{i_{\rm in}}(t)\hat{c}_{i_{\rm out}}}{\Phi_{\rm GS}}}e^{{i\omega t}}dt,
\label{eq:MajoranaG}
\end{align}
where the time-dependent operator is 
{defined by} ${\hat{\mathcal{H}}_{\rm K}}$ 
in which $u_{ij}^\nu$ are 
chosen to realize the flux-free ground state $\ket{\Phi_0}$: 
We take all $u_{ij}^\nu=+1$ for the zigzag case, while we flip $u_{\ij}^\nu$ to $-1$ 
on the $z$ bonds in one column 
for the armchair case {as shown in Fig.~\ref{fig:app_lattice}(a)}. 
In both cases, however, the sign of ${C}_{\rm edge}^{\rm Maj}(\omega)$ is indefinite 
due to the factor of $u_{i_{\rm in} i_{\rm out}}^\nu$; 
we plot the absolute value $|{C}_{\rm edge}^{\rm Maj}(\omega)|$ in Figs.~\ref{fig:hikaku} and \ref{fig:Ldep}, which corresponds to setting 
$u_{i_{\rm in} i_{\rm out}}^\nu = -1$ in Eq.~\eqref{eq:MajoranaG}. 
Note that Eq.~\eqref{eq:MajoranaG} besides this factor corresponds 
to the propagator of the Majorana fermions $\hat{c}_i$.


%

\end{document}